\documentstyle[twocolumn,prl,aps,epsfig]{revtex}

\catcode`@=11
\def\citer{\@ifnextchar [{\@tempswatrue\@citexr}{\@tempswafalse\@citexr[]}}
 
\def\@citexr[#1]#2{\if@filesw\immediate\write\@auxout{\string\citation{#2}}\fi
  \def\@citea{}\@cite{\@for\@citeb:=#2\do
    {\@citea\def\@citea{--\penalty\@m}\@ifundefined
       {b@\@citeb}{{\bf ?}\@warning
       {Citation `\@citeb' on page \thepage \space undefined}}%
\hbox{\csname b@\@citeb\endcsname}}}{#1}}
\catcode`@=12

\def\beq{\begin{equation}}
\def\eeq{\end{equation}}

\def\beqn{\begin{eqnarray}}
\def\eeqn{\end{eqnarray}}
\relax
\def\ba{\begin{array}}
\def\ea{\end{array}}

\newcommand{\lsim}{\raisebox{-0.13cm}{~\shortstack{$<$ \\[-0.07cm] $\sim$}}~}
\newcommand{\gsim}{\raisebox{-0.13cm}{~\shortstack{$>$ \\[-0.07cm] $\sim$}}~}

\begin{document}
\onecolumn
\begin{quote}
\raggedleft
PM/97-36
\end{quote}
\vspace*{3cm}
\begin{center} 
{\bf  Model-independent $\tan \beta$ bounds in the MSSM}
\end{center}
\begin{center} 
Christophe LE MOU\"EL and Gilbert MOULTAKA
\end{center}
\begin{center}
{\it
Physique Math\'ematique et Th\'eorique, UMR-CNRS, \\
Universit\'e de Montpellier II, F--34095 Montpellier Cedex 5, France.}
\end{center}
\date{\today}
\vspace*{3cm}
\begin{center}
{\bf Abstract}
\end{center}
\begin{abstract}
We demonstrate, through the study of
the one-loop effective potential in the MSSM, the existence of fully 
model-independent lower and upper theoretical bounds on $\tan \beta$. 
We give their general analytic form and illustrate some of their implications.
\end{abstract}
\nopagebreak

\twocolumn

\begin{narrowtext}
Phenomenological scenarios for the electroweak symmetry breaking in 
supersymmetric theories have attracted much interest over the past \cite{EWB},
and even more so in the very recent years, motivated by encouraging hints
from accelerator data, like the discovery of a heavy top quark 
\cite{topdiscovery}, 
the possible unification of the three coupling constants \cite{unification}, 
to mention
just two. Several tantalizing theoretical approaches exist of how this
breaking would occur. They are, however, generically tributary of assumptions
about physics at the GUT or Planck scales and thus suffer from some 
related theoretical uncertainties. Whatever the correct theory turns out to be
at these scales though, the low-energy physics 
is thought to be described by
an effective Lagrangian where a linearly realized global supersymmetry
is softly broken and the electroweak symmetry broken at the electroweak
scale. In a typical such model like the Minimal Supersymmetric Standard Model
(MSSM) \cite{susy}, the very many free parameters can be theoretically 
correlated through specific model assumptions together with the requirement 
of radiative electroweak symmetry breaking (EWSB),
leading to phenomenological predictions for the full mass spectrum 
\cite{spectrum}.
In this context, it is commonly assumed that
some theoretical constraints on the effective parameters are obtainable only 
in the above mentioned model-dependent context. 
The aim of the present letter is to investigate a model-independent
alternative which questions this assumption.\\ 
It is indeed important, given the theoretical uncertainties, 
to disentangle the constraints which are
a direct reflection of specific model assumptions from those which proceed 
from general physical requirements. Clearly, the first general physical 
requirement is that the effective potential (of the MSSM) should allow for
a stable EWSB minimum. This requirement should of course be concomitant
with that of an unstable $SU(3) \times SU(2)_L \times U(1)_Y$ gauge invariant
vacuum and of the absence, or at least instability, of color or charge
breaking vacua. [We will have nothing to say about the latter requirement
here and will thus assume for simplicity that it is satisfied\cite{casas}.]  
The conditions for the existence of an electroweak
symmetry breaking minimum at the electroweak scale are usually written
as 
       
\begin{eqnarray}
\frac{1}{2} M_Z^2 = \frac{\overline{m}_1^2 - \overline{m}_2^2 \tan^2 \beta}
{\tan^2 \beta -1 }  &,&
\sin 2\beta= \frac{-2 \overline{m}_3^2}{\overline{m}_1^2 + \overline{m}_2^2}
\label{EWSBcond} \\
\nonumber
\end{eqnarray}   

in the MSSM, where $\tan \beta \equiv \frac{<H_2>}{<H_1>}$ 
is the ratio of the vev's of the two Higgs doublets, and the
$\overline{m}_i$'s involve the soft susy breaking masses and 
are determined from the effective potential (EP).
As they stand, Eqs.(\ref{EWSBcond}) represent a {\sl model-independent}
physical requirement in the sense that they should be fulfilled
irrespective of how the breaking of supersymmetry actually triggers
the EWSB, and whatever is the dynamics underlying supersymmetry breaking 
itself \cite{hiddensector,gaugemed}. We now make an obvious, though crucial, 
observation.
Eqs.(\ref{EWSBcond}) are only first order derivatives, and as such should
not be, generally speaking, expected to define fully a (local) EWSB minimum.
It so happens, however, that they do so in the MSSM but only in the
lowest order of the effective potential, 
namely at tree-level or renormalization-group-improved tree-level
(RGITL) approximations. 
In these approximations the vanishing of the first order derivatives
{\sl implies} the positivity of the second order ones.
[The reader is referred to \cite{nous} for detailed
proofs of this and all subsequent results presented in this letter.]
Therefore, it should come as no surprise that, going beyond these
approximations, one would have to resort to extra conditions from the 
positivity of the second order derivatives and check whether these are
still automatically satisfied. We find that they are generically not, and
imposing them on top of Eqs.(\ref{EWSBcond}) leads to {\sl model-independent}
bounds on $\tan \beta$. A characteristic of the tree-level or RGITL 
approximations
is that the $\overline{m}_i^2$'s in Eqs.(\ref{EWSBcond}) have no dependence on 
$\tan \beta$. If we assume for illustration that $\overline{m}_1^2>0$, 
then one has
the following model-independent constraints:
\begin{itemize}
\item{\sl i)} if $\overline{m}_2^2<0$ then $|\tan \beta| >1$ 
\label{t1}
\item {\sl ii)} if $\overline{m}_2^2>0$ then
$1<|\tan \beta| \leq |\frac{m_1}{m_2}|$ (resp. 
$|\frac{m_1}{m_2}| \leq |\tan \beta|<1$)
for $\overline{m}_1^2 >\overline{m}_2^2$ (resp. 
$\overline{m}_1^2 < \overline{m}_2^2$ )
\label{t2}
\end{itemize}

Let us now go one step beyond the above approximation by considering
the finite (non-logarithmic) one-loop corrections to the effective
potential.  
 
The 1-loop EP \cite{CW} in the $\overline{MS}$ scheme reads 
\begin{equation}
V= V_{tree} + \frac{\hbar}{64 \pi^2} Str[ M^4 (Log \frac{M^2}{\mu_R^2} -
3/2) ] \label{EP}
\end{equation}

where $V_{tree}$ is the tree-level MSSM potential
\cite{susy}, and $M^2$ the field dependent squared mass matrix of the scalar
or vector or fermion 
fields. We will assume for definiteness, hereafter, that all mass scales in 
$M^2$
are of comparable magnitudes. This rough approximation allows a simultaneous 
resummation
of {\sl all} the logs by an appropriate choice of the renormalization scale 
$\mu_R=\mu_R^0$, leading to

\begin{equation}
 V= \overline{V}_{tree}(\mu_R^2) + \frac{\hbar}{64 \pi^2} (-3/2) Str M^4 
\label{Poteff}
\end{equation}

where now $\overline{V}_{tree}(\mu_R^2)$ is obtained from $V_{tree}$ by 
replacing all the tree-level quantities by their running counterparts,
and $Str[...] \equiv \sum_{spin} (-1)^{2 s} (2 s + 1) (...)_s $ sums
over all gauge boson, fermion and scalar contributions. It should be clear
that in the approximation leading to Eq.(\ref{Poteff}) we bypass the
problem of log resummations in the presence of multi-mass scales 
\cite{multiscale}. We consider this as the $0^{th}$ order approximation
in which we state our results, being understood that mass scale disparities
should be ultimately considered. 
Specifying to the Higgs fields directions we find, 

\begin{eqnarray}
&&\overline{V}_{tree}(\mu_R^2) +\kappa Str M^4= \nonumber\\
&&X_{m_1}^2  |H_1|^2 + X_{m_2}^2 
    |H_2|^2 + 
          X_{m_3}^2 (H_1.H_2 + h.c.) + \nonumber\\
         && X ( |H_1|^2 - |H_2|^2)^2 +
        \tilde{\beta} |H_1^{\dagger}  H_2|^2 
+ \tilde{\alpha} ((|H_1|^2)^2 - (|H_2|^2)^2)  
\nonumber\\
\label{vloop}
\end{eqnarray}

where

\begin{eqnarray}
&&\tilde{\alpha} = \frac{3}{2} \kappa g^2({Y_t}^2 - {Y_b}^2) \label{alpha}\\
&&\tilde{\beta}= \frac{\overline{g}_2^2}{2} + \kappa g_2^2 ( g_1^2 + 5 g_2^2 - 
6(Y_t^2 + Y_b^2) ) \\
&&X = \frac{\overline{g}^2}{8} + \kappa (g_1^2 g_2^2  +
   \frac{23 g_1^4 + 5 g_2^4}{4} - \frac{3}{2} g^2({Y_t}^2+ {Y_b}^2 ))  \\
&&X_{m_i}^2 = \overline{m}_i^2 + \kappa  \delta m_i^2 \\
&&\kappa=(-\frac{3}{2})\frac{\hbar}{64\pi^2} \label{kappa}, \;\; 
g^2 \equiv g_1^2 + g_2^2 \\
\nonumber
\end{eqnarray}
where $g_1, g_2$ are respectively the $U(1)_Y, SU(2)_L$ gauge couplings,
$Y_t, Y_b$ the top and bottom Yukawa couplings, 
and $H_1.H_2 \equiv \epsilon_{i j} H_1^i H_2^j$. 
The $\delta m_i^2$'s are functions of the full-fledged MSSM free parameters
and are given elsewhere \cite{nous}. 
The doublets $H_1$, $H_2$ depend each on four real-valued fields
with respect to which one needs to determine the existence and
stability of an EWSB point\footnote{$SU(2)_L$ symmetry allows to gauge
away three fields; we choose, however, to check explicitly the occurrence
of the three goldstone modes}. We also dropped out in Eq.(\ref{vloop})
a cosmological constant which depends on the soft breaking masses
and vanishes with them.
Such a term should be retained in a more refined analysis due to its implicit
dependence on the fields \cite{multiscale}. \\

The effective potential Eq.(\ref{vloop}) has the same functional dependence
as the tree-level, {\sl i.e.} all loop corrections are absorbed in the
definitions of $\tilde{\beta} , X$ and the $X_{m_i}^2$'s [we ignore throughout
the implicit scale dependence on the fields], 
except for the $\tilde{\alpha}$ term
which is a genuine one-loop effect, Eq.(\ref{alpha}). Although quantitatively
small, we will see that this new term changes the qualitative features which
prevailed at the tree-level and RGITL approximations. 

Let us now determine the conditions for the existence of a (local) minimum
which breaks the electroweak symmetry. On one hand,
the eight conditions for a {\sl stationary} point
with respect to the eight real-valued Higgs fields boil down, in the 
neutral direction
\begin{eqnarray} 
<H_1>= \frac{1}{\sqrt{2}}\left( \begin{array}{c} 
               v_1 \\
               0 \\
               \end{array} \right) \hspace{1cm}
<H_2>= \frac{1}{\sqrt{2}}\left( \begin{array}{c} 
               0 \\
               v_2 \\
               \end{array} \right)
\label{neutrdir} 
\end{eqnarray}
($v_1, v_2$ real valued), 
to the two equations: 
\begin{eqnarray}
X_{m_3}^2( \tilde{\alpha} - X) t^4  + ( \tilde{\alpha} X_{m_1}^2 -
X (X_{m_1}^2 + X_{m_2}^2 ) ) t^3 + && \nonumber\\ 
( \tilde{\alpha} X_{m_2}^2 
+ X (X_{m_1}^2 + X_{m_2}^2 ) ) t +  X_{m_3}^2 ( \tilde{\alpha} + X)
=0&& \label{eqtbeta}\\
\nonumber
\end{eqnarray}
\vspace*{-1cm}
\begin{eqnarray}
u = \frac{1}{\tilde{\alpha} (t^2 - 1)}
( X_{m_3}^2(t^2  + 1) + 
(X_{m_1}^2 + X_{m_2}^2 )t ) && \label{eqv1v2} \\
\nonumber
\end{eqnarray}

where $t\equiv \tan \beta$ and $u \equiv v_1 v_2$.
On the other hand, the {\sl stability} conditions at the points satisfying 
Eqs.(\ref{eqtbeta},
\ref{eqv1v2}) give:  

\begin{eqnarray}
&&-(v_1^2 + v_2^2) \frac{X_{m_3}^2}{v_1 v_2} \geq 0 \label{newinv1}
\end{eqnarray}
\vspace*{-.5cm}
\begin{eqnarray}
&&2\tilde{\alpha} ( v_1^2 - v_2^2) + (v_1^2 + v_2^2)(2 X - \frac{X_{m_3}^2}{v_1 v_2} )\geq 0  
\label{newinv2}
\end{eqnarray}
\vspace*{-.5cm}
\begin{eqnarray}
-4 \tilde{\alpha}^2 v_1^2 v_2^2 + 2 (v_2^2 - v_1^2)&&[
(v_1^2 + v_2^2 ) \tilde{\alpha} \nonumber \\
&&- (v_2^2 - v_1^2) X] \frac{X_{m_3}^2}{v_1 v_2}
\geq 0 \label{newinv3}
\end{eqnarray}
\vspace*{-.5cm}
\begin{eqnarray}
( -\frac{X_{m_3}^2}{v_1 v_2} + \frac{\tilde{\beta}}{2}) ( v_1^2 + v_2^2) \geq 0 
\;\; \mbox{(twice)}
\label{newinv4}
\end{eqnarray}
plus three zeros corresponding to the three goldstone degrees of freedom. 
The latter inequalities boil down in turn,
due to the perturbative positivity of $\tilde{\beta}$ and 
$ X \pm \tilde{\alpha}$, to the two conditions 
\begin{equation}
\frac{X_{m_3}^2}{v_1 v_2} \leq 0 \label{newcond1} 
\end{equation}
\begin{equation}
\tan^2 \beta\leq t_{-}\,\,\,
\mbox{or}\,\,\, \geq t_{+}
\label{newcond2}
\end{equation}

where $t_{\pm}$ are defined in Eq.(\ref{tplusminus}) below.

Eqs.(\ref{eqtbeta}, \ref{eqv1v2}) supplemented with the requirement 
$M_Z^2= g^2 u ( t + 1/t)/4$ are a special form of Eqs.(\ref{EWSBcond}),
except that now 
the dependence on $\tan \beta$ in the $\overline{m}_i^2$'s is made 
explicit. They remain fully analytically solvable in our approximation,
but we do not dwell further on this aspect here. Eqs.(\ref{newcond1}, 
\ref{newcond2}) are equivalent to the requirement  that the one-loop
corrected squared Higgs masses be positive, in the approximation
of Eq.(\ref{Poteff}). It is straightforward then to see from 
Eqs.(\ref{newinv1} - \ref{newinv4}) that at tree-level
(or RGITL for that matter), conditions (\ref{newcond1}, \ref{newcond2})
reduce to 
$$ \frac{m_3^2}{v_1 v_2} \leq 0$$
Furthermore, the latter inequality is a direct consequence of 
Eqs.(\ref{EWSBcond}) in this limit, and thus the only necessary bounds
that one can obtain on $\tan \beta$ in a model-independent way are
in this case given by {\sl i)} and {\sl ii)}. 
At the one-loop level the situation becomes more involved. The
requirement that $u$ and $t$ have the same sign ({\sl i.e.}
$v_1$ and $v_2$ are real-valued) together with Eqs.(\ref{eqv1v2}, 
\ref{newcond1}, \ref{newcond2}), allow us to determine new analytic
bounds on $\tan \beta$. Here the sign of $\tilde{\alpha}$ plays
an important role. Furthermore, given its dependence on the
Yukawa couplings, it will concomitantly lead to further consistency
bounds involving $m_t/m_b$, the ratio of the top to down quark masses. 
A detailed analysis \cite{nous} leads to the following bounds 
  
{\bf a)} $\tilde{\alpha} \leq 0$
\begin{eqnarray}
&\mbox{\, if \,} \tan \beta > 1 \;\;\; \mbox{then} \;\;\;
 \max [\sqrt{t_{+}}, T_{+} ]   \leq \tan \beta \leq
\frac{m_t}{m_b}  &\\
\nonumber 
\end{eqnarray}
\vspace*{-1cm}
\begin{eqnarray}
&\!\!\!\!\!\!\!\!\!\!\!\!\!\!\!\!\!\!\!\!\!\!\mbox{\, if \,} \tan \beta < 1 & 
\;\;\; 
\mbox{then} \;\;\; T_{-} \leq \tan\beta \leq \sqrt{t_{-}} \\
\nonumber
\end{eqnarray}

{\bf b)} $\tilde{\alpha} \ge 0$

\begin{eqnarray}
& \max [\frac{m_t}{m_b}, \sqrt{t_{+}} ]\leq \tan 
\beta\leq T_{+}&  \\
\nonumber
\end{eqnarray} 
 
where 
\begin{equation}
T_{\pm} = \frac{ -X_{m_1}^2 - X_{m_2}^2 \mp \sqrt{(X_{m_1}^2 + X_{m_2}^2)^2 -
4  X_{m_3}^4 }}{2 X_{m_3}^2} \nonumber \\
\end{equation}

and
\begin{equation}
t_{\pm}= \frac{ \tilde{\alpha}^2 \frac{v_1 v_2}{X_{m_3}^2} - X  \mp
\sqrt{( X - \tilde{\alpha}^2 \frac{v_1 v_2}{X_{m_3}^2})^2 + \tilde{\alpha}^2-
X^2 }}{ \tilde{\alpha} - X }
\label{tplusminus}
\end{equation}

A couple of remarks are in order. We stress first that the above bounds are
fully model-independent. {\sl No unification or universality assumptions are
needed or assumed}. $T_{\pm}$ and $t_{\pm}$ are generally calculable in
terms of the full-fledged MSSM parameters. When writing the bounds 
{\bf a)} and {\bf b)} we only considered, without loss of generality,
positive $\tan \beta$ values. This is of course just a convention for
the relative phase of the fields $H_1, H_2$. It implies, however, that
$X_{m_3}^2 >0$ is forbidden in view of Eq.(\ref{newcond1}). [An equivalent
discussion can be carried out in the opposite convention.]
From this constraint on the sign of $X_{m_3}^2$ 
and the requirement of boundedness from below of the potential (\ref{vloop})
in the $|H_1|= \pm|H_2|$ direction, namely 
$X_{m_1}^2 + X_{m_2}^2 \pm 2 X_{m_3}^2 \geq 0$, 
it readily follows that $T_{\pm}$ are always real-valued and positive. 
Similarly, $t_{\pm}$  are always real-valued as can be seen from
Eq.(\ref{tplusminus}) upon use of Eq.(\ref{newcond1}), and positive
since $ X \pm \tilde{\alpha} \geq 0$  is always perturbatively satisfied.
Furthermore, one shows that $t_{-} \leq 1 \leq t_{+}$ and
$T_{-} \leq 1 \leq T_{+}$. The inequalities in {\bf a)} and {\bf b)} are
thus always consistent;
In particular, one sees that a band around 
$\tan \beta =1$ {\sl is always excluded}.
More generally, and depending on the chosen values of the MSSM parameters,
the relative magnitudes of $t_{-}, T_{-}$ or $t_{+}, T_{+}$ 
will exclude some domains for $\tan \beta$. 
 
It is also worth noting that the appearance of $m_t/m_b$ in
the bounds is rather peripheral in the sense that it follows from the 
tree-level
Yukawa masses and the form of $\tilde{\alpha}$, not from the
effective potential itself. For instance, had we not neglected the $\tau$
quark Yukawa coupling in $\tilde{\alpha}$, $m_t/m_b$ 
would have been replaced
by $ (m_t/m_b) ( 1 - m_\tau^2/6 m_b^2) $ in {\bf a)} and {\bf b)}. 
Thus the main information 
extracted from (\ref{vloop}) are the model-independent bounds 
$t_{\pm}$ and $T_{\pm}$. It is instructive to compare Eq.(19) with the 
SUGRA-GUT qualitative constraint 
$ 1 \lsim \tan \beta \lsim m_t/m_b$ \cite{giudice}, which relies on 
universality assumptions and the trend of the running of $\overline{m}_1^2$
and  $\overline{m}_2^2$. What we learn here is that the mere consideration of 
the finite (non-logarithmic) contribution to the effective potential improves 
quantitatively the lower bound even if the model assumptions are loosened. 
More generally, conditions {\bf a)} and {\bf b)} distinguish naturally between
small and large $\tan \beta$ and allow to tell when the respective windows are 
closed
or not, due to the general requirement of EWSB. For instance in case {\bf a)}, 
$\tan \beta <1$ would be excluded if $\sqrt{t_{-}} < T_{-}$  while 
$ 1 \lsim \tan \beta \lsim m_t/m_b$  would be forbidden if
$m_t/m_b < \max[\sqrt{t_{+}}, T_{+}]$. Similarly case {\bf b)} would be 
forbidden in the region where $T_{+} <  \max[m_t/m_b, \sqrt{t_{+}}]$.
An exhaustive study of the above situations lies out of the scope of the
present letter as it necessitates a scan over a wide range of the parameter
space. Here we aim at a simple illustration of how conditions 
{\bf a)} and {\bf b)} can be used. For this we chose to correlate the tree-level parts of the $X_{m_i}^2$'s, {\sl i.e.} the $\overline{m}_i^2$'s, by requiring
that they satisfy Eqs.(\ref{EWSBcond}). This eliminates all the Higgs
soft masses (appearing also in the one-loop contributions) in terms of
$\tan \beta^0$ and $m_{A^0}$. The latter variables can be understood as the
guess values for $\tan \beta$ and $m_{A^0}$ (the CP-odd Higgs mass) which
are consistent with EWSB at the {\sl tree-level}\footnote{It should be clear
that such a correlation assumption is just for the sake of illustration
and is by no means mandatory.One could choose the free parameters differently
and compute the bounds from them.}. We then study the behaviour of $T_{+}$
as a function of $\tan \beta^0$ and $m_{A^0}$ after having assigned some values
to the remaining free parameters. Fig.1 illustrates the case {\bf a)}
with $ \tan \beta > 1$ and $ t_{+} < T_{+}$.
\noindent
\begin{figure}[htb]
\begin{center}
\mbox{ 
\psfig{figure=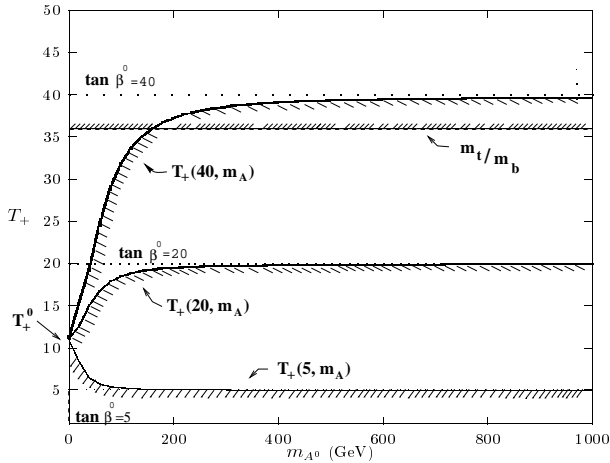,width=13cm,angle=-90}}
\end{center}
\vspace*{-5cm}
\caption[]{$\tilde{\alpha}=-0.0016, Y_t\sim 0.9, A_b=A_t=200,$ $ m_{scalar}=100,
m_{fermion}=300, \mu=-150 (GeV)$}
\end{figure}
 While the usual bounds
\cite{giudice} would have just told us that $\tan \beta^0=5, 20$ are
allowed and $\tan \beta^0=40$ forbidden 
(if we commit ourselves to a given model), we see from Fig.1 that for instance
the tree-level guess $\tan \beta^0=5$ cannot be made consistent with EWSB,
the good candidate values being well above it, unless $m_{A^0} \gsim 100 $ GeV
or so. At the other extreme, $\tan \beta^0= 40$ (a value qualitatively
inconsistent with our input for $Y_t, Y_b$) leads to the situation where
only the region $m_{A^0} \lsim 170$ GeV is allowed and corresponds to a
$\tan \beta$ well below 40. The intermediate guess values lying between
$T_{+}^0 \simeq 11$ and $m_t/m_b$ can be made a priori consistent with
EWSB for any $m_{A^0}$, as seen for instance for $\tan \beta^0 = 20$.
One should keep in mind, however, that we did not require here the symmetry
breaking to be consistent with the Z and top masses yet,
nor did we implement the full information from Eqs.(\ref{eqtbeta},
\ref{eqv1v2}). These would of course constrain further the allowed domains
for $\tan \beta$. It is worth noting that, when $\tilde{\alpha} <<1$, 
$T_{+}$ is a good estimate for
$\tan \beta$ satisfying Eq.(\ref{eqtbeta}). Thus even
when the guess value of $\tan \beta^0$ is not excluded by the lower bound
it remains true, as can be seen for $\tan \beta^0=20$ in Fig.1,
that the correct $\tan \beta$ would appreciably differ 
from it except for very heavy $m_{A^0}$.

To conclude, we believe we have shown that the general form of the
MSSM 1-loop EP in the $\overline{MS}$ scheme contains more information than 
what was {\sl a priori}
expected from model-independent phenomenological analyses. This information
can be easily implemented as analytical constraints on $\tan \beta$ in the MSSM.
Finally, a further treatment of the Logs beyond the naive RGITL 
will certainly improve our approximation, keeping though the above conclusion
qualitatively unchanged.

\end{narrowtext}

\end{document}